\documentclass[lettersize,journal]{IEEEtran}
\usepackage{amsmath,amsfonts}
\usepackage{algorithmic}
\usepackage{algorithm}
\usepackage{array}
\usepackage[caption=false,font=scriptsize,labelfont=sf,textfont=sf]{subfig}
\usepackage{textcomp}
\usepackage{stfloats}
\usepackage{url}
\usepackage{verbatim}
\usepackage{graphicx}
\usepackage{cite}
\usepackage{booktabs}
\usepackage{diagbox}
\usepackage{multirow}
\usepackage{makecell}
\usepackage{hyperref}
\DeclareMathOperator*{\argmin}{arg\,min}
\begin{document}

\title{Non-Negative Matrix Factorization Using Non-Von Neumann Computers}

\author{\IEEEauthorblockN{Ajinkya Borle\IEEEauthorrefmark{1}, Charles Nicholas\IEEEauthorrefmark{2}, Uchenna Chukwu\IEEEauthorrefmark{3}, Mohammad-Ali Miri\IEEEauthorrefmark{4}, and Nicholas Chancellor\IEEEauthorrefmark{5}}\\
\IEEEauthorblockA{\IEEEauthorrefmark{1}\IEEEauthorrefmark{2}CSEE Department, University of Maryland Baltimore County, Baltimore, USA\\
\IEEEauthorrefmark{3}\IEEEauthorrefmark{4}\IEEEauthorrefmark{5}Quantum Computing Inc (QCi), 5 Marine View Plaza, Hoboken,
07030, NJ, USA\\
Email: \IEEEauthorrefmark{1}aborle1@umbc.edu, \IEEEauthorrefmark{3}uchukwu@quantumcomputinginc.com}
}



\maketitle

\begin{abstract}
Non- negative matrix factorization (NMF) is a matrix decomposition problem with applications in unsupervised learning. The general form of this problem (along with many of its variants) is NP-hard in nature. In our work, we explore how this problem could be solved with an energy-based optimization method suitable for certain machines with non-von Neumann architectures. We used the Dirac-3, a device based on the entropy computing paradigm and made by Quantum Computing Inc., to evaluate our approach. Our formulations consist of (i) a quadratic unconstrained binary optimization model (QUBO, suitable for Ising machines) and a quartic formulation that allows for real-valued and integer variables (suitable for machines like the Dirac-3). Although current devices cannot solve large NMF problems, the results of our preliminary experiments are promising enough to warrant further research. For non-negative real matrices, we observed that a fusion approach of first using Dirac-3 and then feeding its results as the initial factor matrices to Scikit-learn's NMF procedure outperforms Scikit-learn's NMF procedure on its own, with default parameters in terms of the error in the reconstructed matrices. For our experiments on non-negative integer matrices, we compared the Dirac-3 device to Google's CP-SAT solver (inside the Or-Tools package) and found that for serial processing, Dirac-3 outperforms CP-SAT in a majority of the cases. We believe that future work in this area might be able to identify domains and variants of the problem where entropy computing (and other non-von Neumann architectures) could offer a clear advantage.

\end{abstract}

\begin{IEEEkeywords}
Matrix decomposition, linear algebra, non-von Neumann architecture, photonics, optimization, metaheuristics, mathematical programming
\end{IEEEkeywords}

\section{Introduction}
\IEEEPARstart{W}{ith} Moore's law slowing down, researchers have been looking into alternative computer architectures to the standard von Neumann computing architecture\footnote{von Nuemann computers are those that store the program and data in the same memory space and adhere to the rules of sequential instruction processing.} \cite{Theis_2017}. These alternative designs are collectively called non-von Neumann (NVN) architectures and can vary significantly in their approach, ranging from application-specific integrated circuits \cite{smith1997application} (mostly implemented using silicon based fabrication technologies) to quantum computing \cite{Nielsen_2012} (with various possible implementation strategies).

Among the various non-von Neumann computers, several have been designed to optimize an energy-based objective function, a task which is NP-hard in general, and return values of the variables that provided optimal results.  These NVN computers include Ising machines that work on binary variables, such as quantum annealing machines \cite{Kadowaki_1998,McGeoch_2014}, probabilistic bit computers \cite{Camsari_2019}, and coherent Ising machines \cite{Wang_2013}, as well as machines that can handle continuous or integer variables such as coherent continuous-value machines \cite{khosravi2022non}, thermodynamic computers\cite{Aifer_2024} and even gate-model quantum computers running continuous quantum approximation \cite{verdon2019quantum}. Problems from different domains, such as machine learning and artificial intelligence, are encoded as the objective function of these machines such that we can derive meaningful results for our original problem from the solutions yielded by the optimization process.

In this work, we propose two different ways to encode and solve non-negative matrix factorization (NMF) problems and evaluate them on the Dirac-3 entropy computer \cite{Nguyen_2025} (implemented with the help of photonics and electronics) from Quantum Computing Inc (QCi). NMF is relevant in the field of unsupervised  machine learning for tasks such as finding a basis of feature vectors that can describe a dataset and dimensionality reduction of the same \cite{Gillis_2020}. Mathematically, this is equivalent of factoring a matrix $V \in \mathbb{R}_{\geq 0}^{n \times m}$ into matrices $W \in \mathbb{R}_{\geq 0}^{n \times p}$ and $H \in \mathbb{R}_{\geq 0}^{p \times m}$ such that $V \approx WH$. Following are the core contributions of our work:
\begin{enumerate}
    \item A formulation that converts the NMF problem into a quartic-degree polynomial (QuarDP) function with continuous or integer variables. This function is not only suitable for entropy computing machines like Dirac-3, but also can, at least in principle, be adapted for NVN solvers such as continuous quantum approximate optimization (CV-QAOA) \cite{verdon2019quantum,Enomoto_2023} and coherent continuous variable machines (CCVM) \cite{khosravi2022non,Brown_2024}.  However, at the time of writing, Dirac-3 is the only continuous variable solver that supports a quartic degree objective function.
    
    \item  A formulation that converts the NMF problem into a quadratic polynomial binary optimization (QUBO) function. This conversion involves (i) approximating continuous and integer variables with binary ones and (ii) reducing the degree of the binary expression from quartic to quadratic by adding addition variables and penalties. It should be noted that this approach to solving the problem can hypothetically run on most types of Ising machines.

    \item A preliminary comparison between our QuarDP approach and the NMF algorithm implemented in the Scikit-learn package~\cite{pedregosa2011scikit}
    for continuous value problems. While the results of using QuarDP on Dirac-3 were worse than Scikit-learn with default parameters, with respect to the relative error in the reconstruction of the original matrix $V$, a fusion of the two approaches generally produced better results than either of the two individual approaches. Our findings suggest that using results of solvers like Dirac-3 as initial matrices $W_{0}$ and $H_{0}$ in Scikit-learn's NMF procedure could be advantageous.

    \item A preliminary comparison between our QuarDP and QUBO approach for solving integer NMF problems. The larger number of variables in the QUBO version of the problem seems to give us a poorer quality result.

    \item A preliminary comparison between our QuarDP approach and the CP-SAT (constraint programming satisfiability) solver for integer matrix problems (for the same amount of time). For the majority of our test cases, we found that while the parallel processing version of CP-SAT outperformed QuarDP on Dirac-3 (based on relative error in reconstructing $V$), the CP-SAT heuristic failed to produce a better result using an individual CPU.
\end{enumerate}
In Section~\ref{sec:related}, we present some background information and a review of related work.
In Section~\ref{sec:formulatingNMF} we describe how we reformulate NMF as an optimization problem.  Section~\ref{sec:experiments} describes the experiments we conducted to support our approach. Finally, we propose some ideas for future work in Section~\ref{sec:futureWork}, and in Section~\ref{sec:conclusion} we present our conclusions.

\section{Background and Related Work}
\label{sec:related}
\subsection{Background}
\subsubsection{Non-negative Matrix Factorization (NMF)}
As mentioned above, the problem of nonnegative matrix factorization (NMF) is one of matrix decomposition where, given a matrix $V \in \mathbb{R}_{\geq 0}^{n \times m}$, the task is to factorize $V$ into matrices $W \in \mathbb{R}_{\geq 0}^{n \times p}$ and $H \in \mathbb{R}_{\geq 0}^{p\times m}$. The goal is to find factors such that $V \approx WH$.

In (unsupervised) machine learning, a dataset can be formatted as a matrix $V$ with $m$ records, each record being a feature vector of length $n$, arranged as the columns of the matrix. 
(This is one of several standard conventions.)
For an integer $p<\min (m,n)$, NMF would produce (i) a matrix $W$ that contains a set of $p$ basis vectors, that is, vectors whose linear combination can recreate the original matrix $V$, and (ii) a dimensionally reduced version of the original dataset in matrix $H$, where each basis vector has a reduced length $p$, where $p < n$.  We assume the reader is familiar with this and other concepts from basic linear algebra.  Dimensionality reduction has been used in a wide variety of domains, such as astronomy \cite{Blanton_2007,Ren_2018}, text mining \cite{Kuang_2014}, computer vision \cite{Guillamet_2002}, and even malware analysis\cite{Eren_2023}, among others.

The general version of the NMF problem is NP-hard \cite{Vavasis_2010} in nature, and most applications use polynomial-time heuristics to get approximations to the solution that correspond to a local minima solution. A principle used in many of these heuristics is block coordinate descent (BCD).
\begin{algorithm}
\caption{Block Coordinate Descent for NMF}
\begin{algorithmic}
\STATE 
\STATE {\textsc{MAIN}}$(\mathbf{V},p,\text{tol},\text{maxcnt})$
\STATE \hspace{0.2cm} Generate initial matrices $W_{0} \in \mathbb{R}_{\geq 0}^{n \times p}$ and $H_{0} \in \mathbb{R}_{\geq 0}^{p \times m}$
\STATE \hspace{0.2cm} Set $i \leftarrow 1$
\STATE \hspace{0.2cm}\textbf{while} $\|V - W_{i}H_{i}\| > \text{tol}$ and $i \leq \text{maxcnt}$
\STATE \hspace{0.4cm} Get $W_i \leftarrow \argmin_{X \in \mathbb{R}_{\geq 0}^{n \times p}} \|V-XH_{i-1}\|$
\STATE \hspace{0.4cm} Get $H_i \leftarrow \argmin_{X \in \mathbb{R}_{\geq 0}^{p \times m}} \|V-W_{i}X\|$
\STATE \hspace{0.2cm} \textbf{end while}
\STATE \hspace{0.2cm} $W \leftarrow W_{i}$,$ H \leftarrow H_i $
\STATE \hspace{0.2cm} \textbf{return} $W,H$
\end{algorithmic}
\label{alg:BCD}
\end{algorithm}

In this form, BCD outlines the general optimization strategy in which the subproblems in it could be solved by different algorithms. The general principle is to iteratively calculate a new matrix $W$ and $H$ until a termination criterion is met. The initialization of the $W$ and $H$ matrices can be done using a wide variety of heuristics, from the creation of matrices with random values, to smarter methods that exploit the principle of singular variable decomposition \cite{Boutsidis_2008}. Similarly, the method by which we calculate $W$ and $H$ is left to us.  One popular method of computing the actual $W_i$ and $H_i$ is by non-negative least squares\cite{Lawson_1995}, although  we note that BCD implemented with non-negative least squares is also known as alternating least squares\cite{Kim_2011} (ALS).

Our method differs from the above by optimizing an objective function based on $\|V-WH\|$ while attempting to find the best values for $W$ and $H$ simultaneously.

\subsubsection{Quadratic Unconstrained Binary Optimization (QUBO) and (Classical) Ising Models}

The Ising model is a mathematical model for ferromagnetism originating in statistical mechanics \cite{Gallavotti_1999}. Although the Ising model was originally invented by physicists in order to better understand ferromagnetism, since the advent of quantum annealers, the model has also gained popularity as a framework through which problems from other domains could be solved using these new quantum annealers.  We note that problems are formulated into an Ising model objective function with the goal of finding variables that optimize the energy function like the one in 
Equation~\ref{eq:ising}. In particular, the objective function of the two-dimensional classical Ising model is given by the following equation:
\begin{align}
    E(\sigma) = \sum_{i} h_{i}\sigma_{i} + \sum_{i<j}J_{ij}\sigma_{i}\sigma_{j}\label{eq:ising}
\end{align}
where $\sigma_{i}\in \{-1,+1\}$ are bipolar variables. $h_{i}$ and $J_{ij}$ are the coefficients for single and quadratic terms, respectively.

The quadratic unconstrained binary optimization \cite{Hammer_1968} (QUBO) is another optimization model, equivalent to Ising, that uses binary variables instead of bipolar ones.
\begin{align}
    F(q) = \sum_{i} u_{i}q_{i} + \sum_{i<j}v_{ij}q_{i}q_{j}\label{eq:qubo} 
\end{align}
where $q_{i} \in \{0,1\}$ are binary variables and $u_i$ and $v_{ij}$ are coefficients for the single and quadratic terms of the expression. The equivalence between Equation~\ref{eq:ising} and Equation~\ref{eq:qubo} is given for each variable $i$ by:
\begin{align}
    \sigma_i &= 2q_{i} - 1\\
    \text{and } F(q) &= E(\sigma) + \text{offset}
\end{align}
Depending on the application, some problems are better suited to the classical Ising model formulation \cite{Kumar_2018,King_2025}, while others work better with QUBO \cite{Date_2021,O_Malley_2018}.
Along with quantum annealers, objective functions like the classical Ising model (and by extension, the QUBO model) have also been adopted by other NVN architecture machines such as entropy computing \cite{Nguyen_2025}, coherent Ising machines \cite{B_hm_2018} and P-bit machines \cite{Camsari_2019}. Devices based on these models are collectively known as Ising machines. It should be noted that although the Dirac-3 can function as an Ising machine, it is not restricted to the Ising/QUBO models, as its primary objective function is more general in nature.
\subsubsection{Entropy Computing and Dirac-3}\label{sec:ec_dirac-3}
Entropy computing (EC) is a photonic-based computing technology that uses an open quantum system to solve non-convex optimization problems by minimizing an objective function. EC is a technology, proposed and currently implemented by Quantum Computing Inc (QCi), which uses temporal photonic modes and photons in those modes to represent qubits and qudits.  (A qudit is a quantum system that stores d-levels of information, where d is an integer greater than 2). Since photonic architectures are expected to consume less power than electronic ones\cite{Xu_2025,Fayza_2025}, entropy computers may offer power consumption advantages over CPUs and GPUs, especially for larger problem sizes.

At this time, the state-of-the-art entropy computer
is the Dirac-3, which implements the EC concept through a photonic-electronic hybrid architecture. See the Appendix for an overview of this device and of entropy computing as a whole. We also recommend the original papers to interested 
readers~\cite{nguyen2024entropy,Nguyen_2025}.

Dirac-3's default objective function is given by
\begin{align}
\begin{split}
    G(x) = &\sum_{a}\mathcal{A}_{a}x_{a} + \sum_{a,b}\mathcal{B}_{ab}x_{a}x_{b} + \sum_{a,b,c}\mathcal{C}_{abc}x_{a}x_{b}x_{c}\\ &+ \sum_{a,b,c,d}\mathcal{D}_{abcd} x_{a}x_{b}x_{c}x_{d}+ \sum_{a,b,c,d,e}\mathcal{E}_{abcde} x_{a}x_{b}x_{c}x_{d}x_{e}\label{eq:dirac-3}
\end{split}  
\end{align}
where $x_i$ can be made to take non-negative real/integer or binary values, or a combination thereof, over a discrete space, in a way that 
depends on how the measured photons are interpreted, the details of which are abstracted from the user. From a user's perspective, an additional parameter $R \in \mathbb{R}$ needs to be set when using Dirac-3 for optimization problems with non-negative real values. The sum of all variable values must be equal to $R$. In other words,
\begin{align}
    \sum_{i} x_i = R\label{eq:dirac-3constraint}
\end{align}

This possible values of $x_i$ define the interval $[0,R]$.
Given the device's maximum resolution of 10,000 discrete levels, this interval is partitioned into ten thousand equally-spaced values. For certain applications, an extra (slack) variable multiplied explicitly by a zero coefficient $0x_s$ could be used as part of the objective function. The purpose of this slack variable is to absorb a portion of $R$ that is not used by the main objective function while still satisfying the summation constraint of Equation~\ref{eq:dirac-3constraint}.
The device can also function as an integer or binary solver. In this mode, the user programmatically sets the number of discrete levels per variable, where a variable with $n$ levels
can take integer values in $\{0, 1, \ldots, n-1\}$ with a total of 954 levels available collectively.
In this configuration, the user does not need to set the value of $R$, and every variable $x_i$ can have a different number of discrete values.

\subsection{Related Work}
In the area of applying NVN architecture computers to the problem of matrix factorization, we select two important works for special mention. In the first paper, Kerenidis and Prakash propose a method for singular value decomposition (SVD) on universal quantum computers \cite{kerenidis_2017}. In the second paper, O’Malley et al. use quantum annealers to do non-negative/binary matrix factorization (NBMF), a variant of NMF where the matrix $H$ is binary~\cite{O_Malley_2018}. The latter is more relevant to our work as it (i) works on a variant of NMF rather than SVD and (ii) uses a QUBO formulation. However, unlike their method, ours (i) works on the standard and integer variants of NMF, and (ii) computes the factor matrices $W$ and $H$ simultaneously while theirs use an ALS scheme. 

Ottaviani and Amendola~\cite{ottaviani2018low} use a D-wave quantum annealer to solve low-rank NMF problems. They proposed using multiple qubits together to approximate individual variables with integer or real values, but their approach still employs the alternating scheme from the BCD/ALS methodology. In 2021, Malik et al.~\cite{Malik_2021} described how a digital annealer (an Ising machine) could be used to factorize binary matrices. Unlike alternating optimization methods, their approach calculated both factor matrices simultaneously, which we do also. Finally, recent works have implemented NBMF with the reverse annealing features of D-wave's quantum annealers \cite{Golden_2021,Haba_2025}.

\section{Formulating NMF as an Energy-Based Optimization Problem}
\label{sec:formulatingNMF}
\subsection{The core idea}\label{sec:nmf_obj_func_dirac-3}
Taking inspiration from Algorithm~\ref{alg:BCD}, we begin with the matrix norm of the difference between $V$ and $WH$
\begin{align}
    \arg \min_{W,H}\|V - WH\| \label{eq:nmf_norm_basic}
\end{align}
The goal is to find matrices $W$ and $H$ such that $\|V-WH\|$ is minimized. 

In Equation~\ref{eq:nmf_norm_basic}, if we specify the Frobenius norm, we have the following:
\begin{align}
 \|V-WH\|_{F} = \sqrt{\sum_{i=1}^{n}\sum_{j=1}^{m}|V_{ij} - \sum_{k=1}^{p}W_{ik}H_{kj}| ^2}\label{eq:nmf_frobnorm}
\end{align}
Taking the square of 
Equation~\ref{eq:nmf_frobnorm}, we get
\begin{align}
\|V-WH\|_{F}^{2} = \sum_{i=1}^{n}\sum_{j=1}^{m}|V_{ij} - \sum_{k=1}^{p}W_{ik}H_{kj}|^2 \label{eq:nmf_frobnorm_sq_basic}
\end{align}
Finally, since none of the values of $V,W$ or $H$ are complex, we can drop the absolute value operation since $|.|^2 = (.)^2$.
\begin{align}
\|V-WH\|_{F}^{2} = \sum_{i=1}^{n}\sum_{j=1}^{m}(V_{ij} - \sum_{k=1}^{p}W_{ik}H_{kj})^2 \label{eq:nmf_frobnorm_sq}
\end{align}
Expanding Equation~\ref{eq:nmf_frobnorm_sq}, we get
\begin{align}
\begin{split}
\|V-WH\|_{F}^{2} &= \sum_{i=1}^{n}\sum_{j=1}^{m}(V_{ij})^{2} + \sum_{k=1}^{p}(W_{ik}H_{kj})^2\\ &- 2V_{ij}\sum_{k=1}^{p}W_{ik}H_{kj}\\ &-2\sum_{k=1}^{p}\sum_{l=1,l\neq k }^{p} W_{ik}W_{il}H_{kj}H_{lj}\label{eq:nmf_frobnorm_sqfinal}
\end{split}
\end{align}
Equation~\ref{eq:nmf_frobnorm_sqfinal} is the essence of our quartic degree polynomial (QuarDP) objective function. For Dirac-3, the variables in $x$ are assigned to the individual elements in $W$ and $H$. Ignoring the terms that only contain constants, such as $\sum_{i=1}^{n}\sum_{j=1}^{m}V_{ij}$, the other terms of Equation~\ref{eq:nmf_frobnorm_sqfinal} can be mapped to the quadratic and quartic terms of Equation~\ref{eq:dirac-3}. 
Finally, the value of $R$ in Equation~\ref{eq:dirac-3constraint} can either be user-defined or chosen algorithmically, like a hyperparameter. With this information, Dirac-3 attempts to find values for $W$ and $H$ that minimize the objective function. To make sure that the sum constraint does not interfere with Dirac-3's ability to minimize the main objective function, we also introduce a slack variable $x_s$ multiplied with a zero coefficient, as mentioned above.

This quartic polynominal approach can also work for other NVN machines that operate on energy-based objective functions capable of supporting quartic terms, such as CCVM \cite{verdon2019quantum} and CV-QAOA \cite{verdon2019quantum,Stein_2023}, albeit with suitable preprocessing for compatibility.

\subsection{The QUBO Formulation}
While Equation~\ref{eq:nmf_frobnorm_sqfinal} can be used to map an integer or real-valued problem directly to an NVN machine like Dirac-3, this subsection describes the process for creating a QUBO objective function suitable for Ising machines. The two main processes involved in doing so are (i) converting real and integer valued variables into binary variables and (ii) converting the quartic objective function into a quadratic one.

\subsubsection{Converting Real and Integer Variables to Binary}

O'Malley and Vesselinov \cite{O_Malley_2016} were the first to suggest a way to approximate real and integer variables as binary variables. If we take a variable $x_{i} \in \mathbb{R}$, a binary approximation can be done by the following
\begin{align}
    x_{i} =  \kappa \sum_{j=0}^{N}2^{j}q_{i}^{(j)} + C\label{eq:binarization}
\end{align}
That is, each variable $x_{i}$ is represented by $N+1$ binary variables $q_{i}^{(j)}$ ($0\leq j \leq N$), a scaling factor $\kappa \in \mathbb{R}$, and a constant offset $C\in \mathbb{R}$. 
Equation~\ref{eq:binarization} can allow for a discretization of different continuous intervals depending on the choice of $N,\kappa$ and $C$. For example, $N=3,\kappa=1$ and $C=-4$ gives us all nine integers in the interval $[-4,4]$.

Using the technique from Equation~\ref{eq:binarization} on the $W$ and $H$ variables in Equation~\ref{eq:nmf_frobnorm_sq}, we get
\begin{align}
\begin{split}
    \|V-W\|_{F}^{2} = &\sum_{i=1}^{n}\sum_{j=1}^{m}(V_{ij} - \sum_{k=1}^{p}(\kappa \sum_{l=0}^{N}2^{l}q_{W_{ik}}^{(l)} + C)(\kappa\\ &\sum_{l=0}^{N}2^{l}q_{H_{kj}}^{(l)} + C))^2\label{eq:nmf_frobnorm_sq_binary}
\end{split}
\end{align}

Since we are dealing with binary terms, the binary variables $q_i$ would be idempotent under multiplication \cite{rosen1999discrete}, such that $(q_{i}^{(j)})^{z} = (q_{i}^{(j)})^{1}$ where $z\in\mathbb{Z}_{>0}$. The appropriate substitutions with respect to idempotence of these variables must be made after 
Equation~\ref{eq:nmf_frobnorm_sq_binary} is expanded. For doing this task in programmatic fashion, we recommend using a symbolic computing package (such as {\tt sympy} in Python) so that you do not need to track every single term  of the objective function manually.

\subsubsection{Converting the Binary Problem from Quartic to Quadratic}\label{sec:quartic_2_quadratic}
To transform the expanded expression from Equation~\ref{eq:nmf_frobnorm_sq_binary} into quadratic form, we will apply insights from research on reducing pseudo-boolean polynomial expressions in optimization problems \cite{rosenberg1975reduction,dattani2019quadratization}. A common approach is to use auxiliary variables and penalty terms to ensure the main and auxiliary variables work as desired. To take a trivial example, suppose we need to minimize the following function:
\begin{align}
    f_{1}(q) = a(q_{1}q_{2}q_{3}q_{4})\label{eq:q1q2q3q4}
\end{align}
where $a\in\mathbb{R}$ is a coefficient. We introduce two new variables $y_{1}$ and $y_{2}$ such that
\begin{align}
    y_{1} = x_{1}x_{2}\label{eq:y1}\\
    y_{2} = x_{3}x_{4}\label{eq:y2}
\end{align}
An Ising machine cannot enforce the equalities of Equations~\ref{eq:y1} and~\ref{eq:y2} directly, so we need to incorporate them into the function to be minimized.
\begin{align}
    \begin{split}
    &f_2(q,y) \equiv f_1(q) = a(y_1y_2) + \lambda(3y_1 + 3y_2 + q_1q_2 + q_3q_4\\ 
    &-2(q_1y_1 + q_2y_1) - 2(q_3y_2 + q_4y_2))
    \end{split}\label{eq:q1q2q3q4_quad} 
\end{align}
With the help of a penalty value $\lambda\in \mathbb{R}_{>0}$, we arrive at Equation~\ref{eq:q1q2q3q4_quad}, a quadratic expression that is designed to be functionally equivalent to Equation~\ref{eq:q1q2q3q4}. 

A similar approach was used by Malik et al. for binary matrix factorization on digital annealers~ \cite{Malik_2021}. In our work, we used the {\tt make\_quadratic()} functionality in D-wave's Ocean SDK for quadratization.

\section{Experiments}
\label{sec:experiments}
Although future machines are expected to scale up the number of variables for quartic terms \cite{Nguyen_2025}, at this point the Dirac-3 machine can have up to 39 variables for a quartic objective function~\cite{qci_dirac3_2024}. Therefore, in the preliminary experiments we now describe, we are limited to 39 variables. The dimensions $p,n$ and $m$ must therefore adhere to
\begin{align}
    p(n+m) + 1 \leq 39 \label{eq:dirac-3_limit}
\end{align}
Here, $p(n+m)$ is the total number of elements of the factor matrices $W$ and $H$ combined, and an additional variable is used as the slack variable as mentioned in Section~\ref{sec:ec_dirac-3}. Each test case in all our experiments was run 10 times. 

All our experiments were developed in the Python programming language on an ordinary laptop with an Intel\textsuperscript{TM} i5-1335U processor and 32 GB of RAM. We used the {\tt qci\_client} package from QCi to communicate with a Dirac-3 machine on QCi's cloud platform. Two other notable packages were used in these experiments: {\tt Scikit-learn v1.3.1}~\cite{pedregosa2011scikit} for NMF with real-valued matrices (henceforth referred to as sklearn), and {\tt Or-Tools v9.14.6206}~\cite{ortools} (henceforth referred to as ORTOOLS) for NMF problems with integer matrices. Our use of a mere laptop when comparing against Dirac-3 is justified because (i) the problem sizes we are dealing with are relatively small, and do not need desktop/server CPUs, and (ii) the i5-1335U is representative of mid-range laptop CPUs as of the time of writing~\cite{intel_i5_1335u_specs} with 12 logical processors (henceforth referred to as just `processors') for parallel processing, providing an adequate amount of computational power, and (iii) the technology for Dirac-3 is still in its nascent stages such that a comparison against more powerful desktop/server CPUs would not be equitable. 

The test cases for real NMF all have factor matrices $W\in [0,1)^{n\times p}, H\in[0,1)^{p\times m}$ in order to better work with Dirac-3. This is not an unrealistic criterion, since for a real-world problem we can scale the values of any arbitrary non-negative real matrix to $V\in [0,1)^{n\times m}$.
Note that $W\in [0,1)^{n\times p}, H\in[0,1)^{p\times m}$ does not imply $V\in [0,1)^{n\times m}$. But  $V\in [0,1)^{n\times m}$ does imply $W\in [0,1)^{n\times p}, H\in[0,1)^{p\times m}$. 

\subsection{Experiment I}\label{sec:exp1}
\subsubsection{Setup}\label{sec:exp1_setup}

In this first experiment, we want to directly compare the results from Dirac-3 and the conventional solver from {\tt Scikit-learn}'s NMF procedure. The latter uses the fast hierarchical ALS technique (fast HALS) by 
default~\cite{CICHOCKI_2009} 
which works on an objective function based on minimizing $\|V-WH\|_{F}$.

For this experiment, we generate test cases where matrix $V$ is created by multiplying randomly generated factor matrices $W$ and $H$, both generated using the {\tt rand()} function from the {\tt numpy} package and then rounding it to the hundredth place. This was done for square matrices of size $1\times1,2\times2,3\times3$ and $4\times4$. We could not go to $5\times 5$ or higher as that would involve more than 39 variables. 
For now, we set $n=m=p$. Although cases with $n=m=p= 1$ essentially involve factorizing scalars, they remain important for gauging the performance of the Dirac-3 device.  A total of 100 test cases were generated for each size. Although square-to-square NMF problems, where $V, W, H \in \mathbb{R}^{n\times n}$ have limited practical applications, this experiment aims to provide a preliminary performance analysis of matrix factorization algorithms on Dirac-3 as the dimensionality increases.

On Dirac-3, each test case was encoded using the formulation mentioned in 
Section~\ref{sec:nmf_obj_func_dirac-3}. The test case was then sent to the machine to be processed with relaxation schedules
1,2 and 3. 
The relaxation schedule refers to the optimization strategy by which the entropy computer adjusts its internal parameters over time as it seeks a global minimum. Of the three, relaxation schedule 1 is the fastest but the least thorough, while relaxation schedule 3 is the opposite.
The value of $R$ was taken to be $n\times p + p\times m = p(n+m) = 2n^2$ as  $2n^2$ is  the count of all the values in $W$ and $H$ combined, and 
recalling that no value in $W$ and $H$ could be greater than one. The run time of Dirac-3 is decided by the machine internally according to the relaxation schedule chosen and the size of the problem submitted.

Upon getting the results back from the Dirac-3 machine, the results were mapped to their respective values in $W$ and $H$. We then reconstructed the matrix $V$ as $\tilde{V} = WH$. To assess the quality of the factorization, we use the absolute and relative error metrics defined by
\begin{align}
    \epsilon_{V} &= \|V-\tilde{V}\|_F\label{eq:abs_error}\\
    \delta_{V} &= \frac{\epsilon_{V}}{\|V\|_F} = \frac{\|V-\tilde{V}\|_F}{\|V\|_F}\label{eq:rel_error}
\end{align}
Where the goal is to have error values as close to 0 as possible. Equations~\ref{eq:abs_error} 
and~\ref{eq:rel_error} use the Frobenius norm for the sake of consistency with our objective function being based on the same norm.

We compared the results from Dirac-3, selecting the run with the lowest relative error, with the results obtained from {\tt Scikit-learn}'s NMF implementation. For the latter, we used the default parameter settings provided by the package, including the popular NNDSVDA (Non-negative double singular value decomposition with averaging) algorithm \cite{Boutsidis_2008} for setting the initial matrix factors $W_{0}$ and $H_{0}$. The default parameters used in Scikit-learn models are established through rigorous evaluation by domain experts and represent well-validated baselines for comparative analysis \cite{pedregosa2011scikit}, making these the most appropriate benchmarks available to us for our evaluation.

\subsubsection{Results and Discussion}\label{sec:exp1_results}
\begin{figure}[htbp]
    \centering
    \subfloat[]{
        \includegraphics[width=8cm, height=6cm]{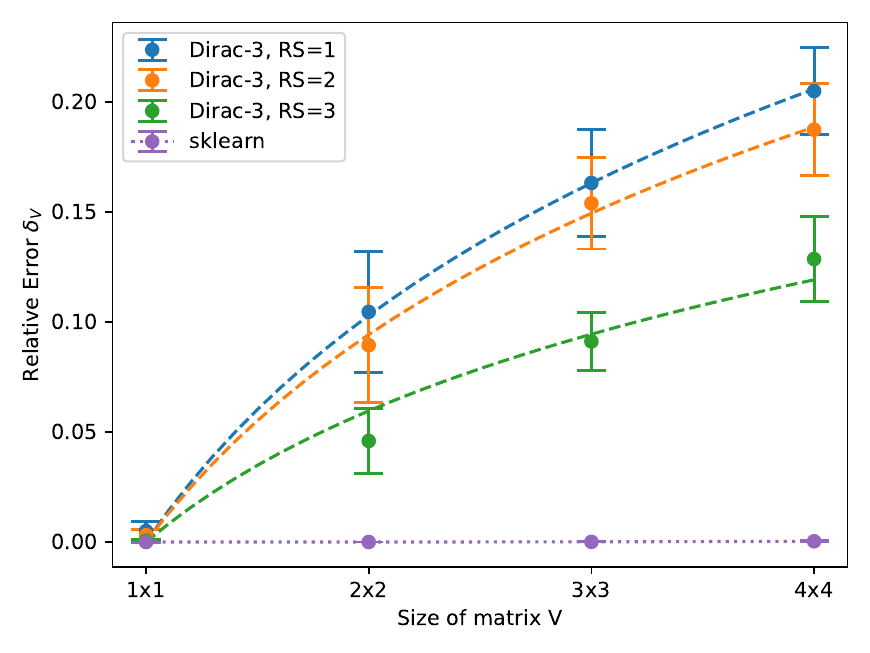}
        \label{subfig:exp1_rel_error}
    }
    \hfill
    \subfloat[]{
        \includegraphics[width=8cm, height=6cm]{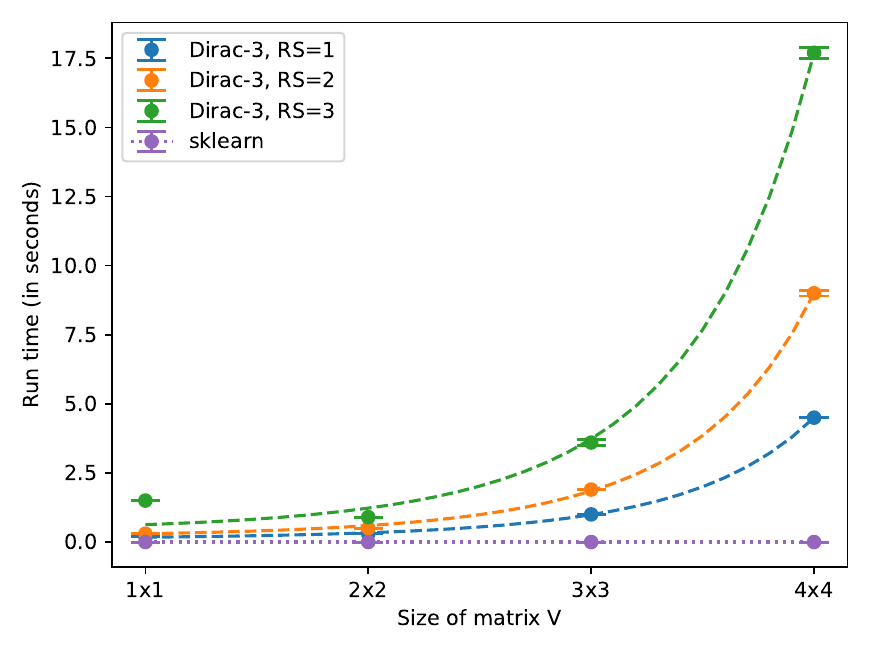}
        \label{subfig:exp1_time}
    }
    \caption{(a) Median relative errors $\delta_{V}$ of the reconstructed matrices $\tilde{V}$ for the test cases in experiment I (with respect to their size) and (b) Median runtime in seconds for the same (for one run). The Dirac-3 results correspond to their relaxation schedules (RS) for which they were run, and sklearn refers to the results from Scikit-learn. The dashed lines for the results of Dirac-3 are curves fitted to the corresponding medians.}
    \label{fig:exp1}
\end{figure}

Figure \ref{fig:exp1} shows the growth of the relative errors $\delta_{V}$ and runtime (in seconds) as a function of the size of the problem, i.e. the size of matrix $V$.
The median relative error  of all the test cases for a given problem size is plotted as a data point in Figure~\ref{subfig:exp1_rel_error}, and the corresponding error bars show the median absolute deviation (MAD) for that median. Similarly, 
Figure~\ref{subfig:exp1_time} shows the median and MAD of the run time in seconds for the same.  We did not set Dirac-3's run time: it was set by the {\tt qci\_client} package based on the relaxation schedule and the size of the problem. Curve fitting was done for the median data points from Dirac-3's results for different relaxation schedules RS. We fitted logarithmic curves to relative error growth and exponential curves to run time growth, respectively. While logistic curves provided a better statistical fit for run time, we chose exponential curves as they are more theoretically consistent with NP-hard computational complexity.

This preliminary experiment appears to suggest that when compared directly against industry standard tools like Scikit learn, Dirac-3 returns lower quality results, in terms of $\delta_{V}$, for higher computational run times, at least for small problems. Here, it should be noted that (i) sklearn is a very sophisticated tool based on decades of research, (ii) Dirac-3's precision for representing variables with real values, while impressive for a nascent technology, is no match for the floating point capability of a traditional computer, and (iii) it is quite possible that Dirac-3 may still prove useful for NMF when used in a different setting (see Section \ref{sec:exp2}), or for different variants of NMF (see Section~\ref{sec:exp4}).


\subsection{Experiment II}\label{sec:exp2}
\subsubsection{Setup}\label{sec:exp2_setup}
For this second experiment, we generate matrices $V\in\mathbb{R}^{4\times8}$ for which we seek to find factor matrices $W\in \mathbb{R}^{4\times3}$ and $H\in\mathbb{R}^{3\times8}$. These problems are more representative of real-world NMF applications as they involve rectangular factor matrices with $p < \min(n,m)$. Two sets of test cases were generated as follows:
\begin{enumerate}
    \item Test set A: One hundred test cases where, for each test case, the matrix $V\in\mathbb{R}^{4\times8}$ is generated using the {\tt rand()} function in numpy and then rounded to the hundredth place.
    \item Test set B: One hundred test cases where, for each test case, we first generate factor matrices $W\in \mathbb{R}^{4\times3}$ and $H\in\mathbb{R}^{3\times8}$ (both generated using the {\tt rand()} function from the {\tt numpy} package and then rounded to the hundredth place). The matrix $V$ is then generated by multiplying $W$ and $H$.
\end{enumerate}
Test set B represents datasets that are inherently amenable to factorization; these matrices are guaranteed to have an exact factorization since they are constructed as $V = WH$. In contrast, test set A represents datasets for which there is no guarantee that an exact decomposition exists. The dimensions of $n,p$ and $m$ were chosen to create problems that were close to the maximum number of variables that Dirac-3 could process for quartic equations. Substituting $n=4,p=3$ and $m=8$ in Equation~\ref{eq:dirac-3_limit}), we can see that the inequality holds and the total number of variables for each of these problems in Dirac-3 is 37.
We ran each problem 10 times on Dirac-3 using relaxation schedules 1 and 2.  Relaxation schedule 3 was not chosen because of the disproportionately long runtimes for our problem size.

As in Experiment I, problems from both test sets were run using sklearn's NMF procedure with default parameters. Initial testing, based on that first experiment, made us doubt that Dirac-3's raw performance would be any better than that of sklearn's. So we also ran the problems using the combination of Dirac-3's results and sklearn's NMF. This entailed the following steps for a given problem:
\begin{enumerate}
    \item From the results of the ten runs of Dirac-3,for all the schedules involved, we compute the median of their energies (or objective function values) and select the result with the energy closest to the median.
    \item From the chosen result, we decode the factor matrices $W^{(Dirac-3)}$ and $H^{(Dirac-3)}$ to be used as the initial matrices in sklearn.
    \item We use sklearn on the problem with default parameter settings, but with the initial matrices given by $W^{(Dirac-3)}\rightarrow W_{0}$ and $H^{(Dirac-3)}\rightarrow H_{0}$.
\end{enumerate}
In a nutshell, we are trying to assess what if any value Dirac-3 and entropy computing could add by creating better initial factor matrices. Let $W^{(fusion)},H^{(fusion)}$ be the factor matrices we get from the fusion approach and $W^{(sklearn)},H^{(sklearn)}$ denote those obtained using sklearn alone. Using the factors for each of these approaches, we reconstruct matrix $V$  as $\tilde{V}^{(fusion)}$ and $\tilde{V}^{(sklearn)}$ respectively.
\subsubsection{Results and Discussion}\label{sec:exp2_results}
\begin{figure*}[t]
\centering
\subfloat[]{\includegraphics[width=8cm, height=6cm]{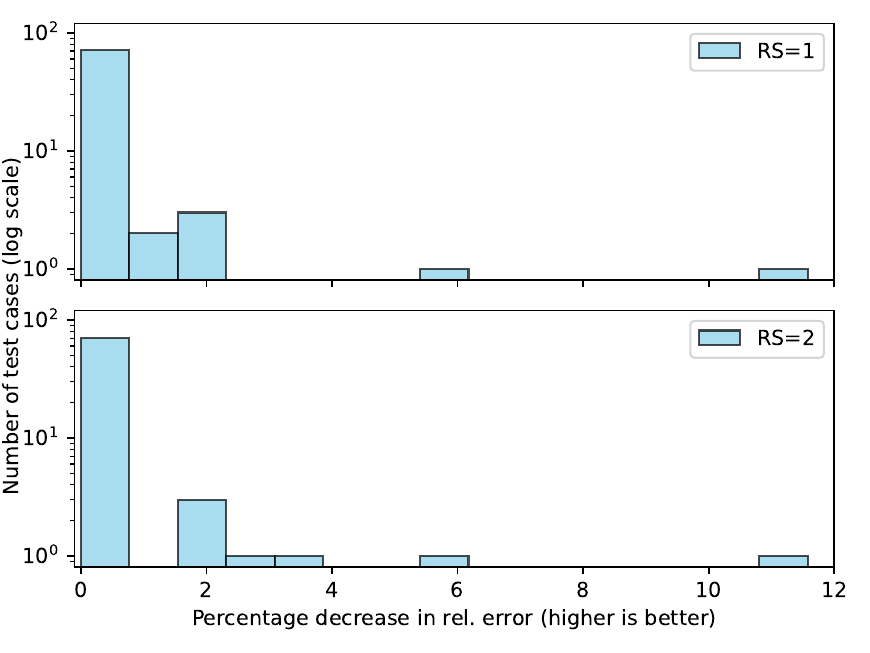}%
\label{subfig:hist_exp2_setA}}
\hfil
\subfloat[]{\includegraphics[width=8cm, height=6cm]{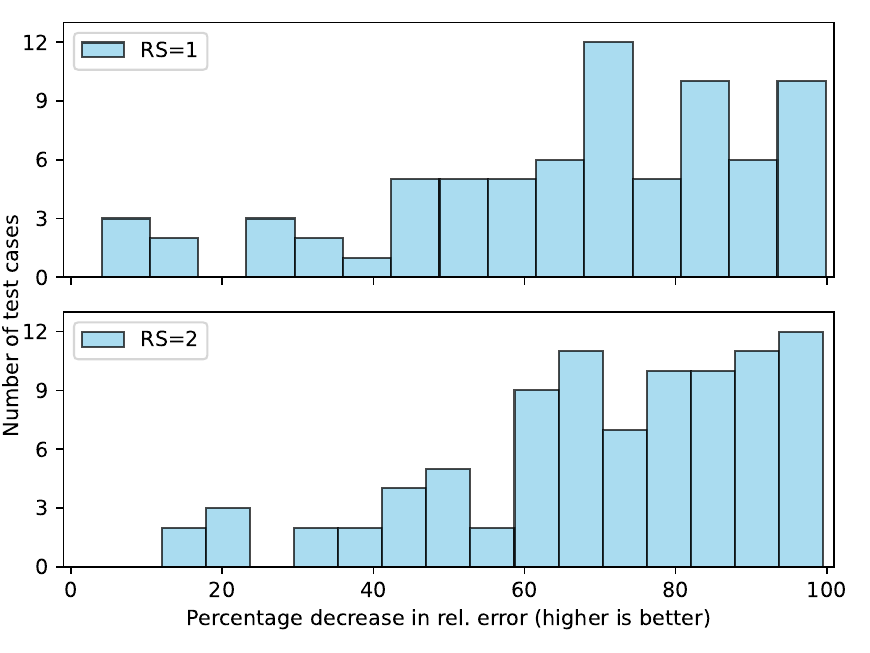}%
\label{subfig:hist_exp2_setB}}
\caption{Histograms for the test cases where the fusion approach outperformed the plain sklearn method for (a) test set A and (b) test set B respectively as a measure of the percentage improvement in relative error $\Delta\delta_{V}$ (RS refers to the relaxation schedule of Dirac-3). Note that the axes in the sub-figures have different scales, to appropriately display the range of results for each test set.} 
\label{fig:hist_exp2}
\end{figure*}

As we said in Section~\ref{sec:exp2_setup}, initial testing revealed that the matrix reconstructed from the output of Dirac-3 $\tilde{V}^{(dirac3)} \leftarrow W^{(dirac3)}H^{(dirac3)}$ was always inferior, in terms of relative error, to that produced with the sklearn approach $\tilde{V}^{(sklearn)} \leftarrow W^{(sklearn)}H^{(sklearn)}$.

However, when comparing $\delta_{V}^{(fusion)}$ and $\delta_{V}^{(sklearn)}$, computed using $\tilde{V}^{(fusion)}$ and $\tilde{V}^{(sklearn)}$) in Equation~\ref{eq:rel_error} respectively, we find that there are many test cases where the fusion approach gives a better result than using sklearn's approach alone. Tables \ref{tab:which_better_exp2_setA} and \ref{tab:which_better_exp2_setB} show that for the majority of cases, for both relaxation schedules (labeled RS), the fusion approach outperforms the plain sklearn method in terms of relative error. To verify the statistical significance of our results, we also computed the p-value from the counts using the binomial test \cite{Lehmann_1993}.
\begin{align}
p = \frac{1}{2^{n_b + n_w}} \sum_{k=n_b}^{n_b + n_w} \binom{n_b + n_w}{k}\label{eq:p-val}
\end{align}

In this one-sided test, the null-hypothesis $H_{0}$ is that the fusion approach is not better than the plain sklearn approach, with respect to relative errors, for the problems they are tested on. The alternative hypothesis $H_{1}$ is that the fusion approach is better than sklearn's. We compute p-values from the  perspective of an expected winner, in this case, the fusion approach. 
In Equation~\ref{eq:p-val}, $n_b$ is the number of test cases where the expected winner did better and $n_w$ is the number of test cases where it did worse.
As is the convention in statistics, we shall take $p<0.05$ to indicate statistically significant support for the alternative hypothesis $H_{1}$.
From Tables \ref{tab:which_better_exp2_setA} and \ref{tab:which_better_exp2_setB}, we see that the p-values overwhelmingly support our alternative hypothesis.

For test set A, moving from RS 1 to RS 2 does not increase the number of cases where the fusion approach performs better, but it does increase them for test set B. This may be because the cases in the latter set have at least one pair of factor matrices, making it possible for the relative error to be closer to 0 than for test set A. We must mention, however, that it is too early to characterize the effect of relaxation schedules using just this one result. For that, further research may be required.

To better understand the quality of the results,
we go on to use the metric of percentage decrease for relative errors defined as
\begin{align}
    \Delta\delta_{V} = \frac{\delta_{V}^{(sklearn)} - \delta_{V}^{(fusion)}}{\delta_{V}^{(sklearn)}} \times 100\label{eq:rel_error_percentage}
\end{align}
From Equation~\ref{eq:rel_error_percentage}, we can see that $\Delta\delta_{V} \in (-\infty,100]$. A case where $\Delta\delta_{V}$ is negative implies a percentage increase rather than a percentage decrease, the value for which being $|\Delta\delta_{V}|$). Using this metric, we create 
Tables~\ref{tab:percentage_change_setA} and \ref{tab:percentage_change_setB}. From these tables, we can see that the fusion results appear to be better for test set B than for test set A. In fact, while the median improvement for test set B is between 70-75\%, the median improvement for test set A is close to zero. Furthermore, going from relaxation schedule 1 to 2 also does not help in test set A. These results, although not yet subjected to statistical tests, indicate that there may be a larger scope for improvement in cases where there exists one or more matrix pairs $W,H$ such that $\|V-WH\|$ is small, if not 0. The other factors that may have influenced these results are the default parameters of sklearn (the justification for which was given in 
Section~\ref{sec:exp1_setup}), and changing them may produce different results.

\begin{table}[htbp]
    \centering
    \begin{tabular}{|c|c|c|}
    \hline
        \diagbox{\textbf{Method}}{\textbf{RS (for fusion)}} & \textbf{RS=1}& \textbf{RS=2} \\
        \hline
         fusion better count & 78 & 77 \\
         \hline
         sklearn better count & 22 & 23\\
         \hline
         p-value (fusion better) & 7.95e-9  & 2.75e-8\\
         \hline
    \end{tabular}
    \caption{Performance comparison by lower relative error: fusion vs. sklearn methods (test set A)}
    \label{tab:which_better_exp2_setA}
\end{table}

\begin{table}[htbp]
    \centering
    \begin{tabular}{|c|c|c|}
    \hline
        \diagbox{\textbf{Method}}{\textbf{RS (for fusion)}} & \textbf{RS=1}& \textbf{RS=2} \\
        \hline
         fusion better count & 75 & 90 \\
         \hline
         sklearn better count & 25 & 10\\
         \hline
         p-value (fusion better) & 2.82e-7 & 1.53e-17\\
         \hline
    \end{tabular}
    \caption{Performance comparison by lower relative error: fusion vs. sklearn methods (test set B)}
    \label{tab:which_better_exp2_setB}
\end{table}

\begin{table}[htbp]
\centering
\begin{tabular}{|l|c|c|}
\hline
\textbf{Performance Measure} & \textbf{Schedule 1} & \textbf{Schedule 2} \\
\hline
Best Improvement & 11.5768\% & 11.5768\% \\
\hline
Worst Change & 9.2894\% increase & 9.7023\% increase \\
\hline
Median Improvement Value &  0.0001 \% & 0.0001\% \\
\hline
MAD Improvement Value & 0.0001 & 0.0001 \\
\hline
\end{tabular}
\caption{Percentage relative error changes: fusion over Sklearn (test set A)}
\label{tab:percentage_change_setA}
\end{table}

\begin{table}[htbp]
\centering
\begin{tabular}{|l|c|c|}
\hline
\textbf{Performance Measure} & \textbf{Schedule 1} & \textbf{Schedule 2} \\
\hline
Best Improvement & 99.8797\% & 99.5046\% \\
\hline
Worst Change & 2706.4505\% increase & 507.4047\% increase \\
\hline
Median Improvement Value & 70.3181\% & 75.1576\% \\
\hline
MAD Improvement Value & 14.8157 & 12.7089 \\
\hline
\end{tabular}
\caption{Percentage relative error of fusion Changes Over Sklearn (test set B)}
\label{tab:percentage_change_setB}
\end{table}

Finally, the histograms in Figure~\ref{fig:hist_exp2} illustrate the spread of $\Delta\delta_{V}$ for the two test sets, particularly for the cases where the fusion results were better. Figure~\ref{fig:hist_exp2}(a) uses a logarithmic scale (base 10) on the Y-axis because most test cases in test set A had $\Delta\delta_{V}\in [0,1]$, creating a highly skewed distribution. Figure~\ref{fig:hist_exp2}(b) shows the distribution skewed slightly to the right for $RS=2$ as compared to for $RS=1$. This also happens in Figure~\ref{fig:hist_exp2}(a), but by a very small margin.

\subsection{Experiment III}\label{sec:exp3}
\subsubsection{Setup}\label{sec:exp3_setup}
For this third experiment, $V, W$ and $H$ are all integer matrices. We created 100 test cases with square matrices $W,H\in \{i\in\mathbb{Z} | 0 \leq i \leq 7\}^{n\times n}$ per $n \in \{1,2,3,4\}$ from which $V \leftarrow WH$ was generated. The goal was to make a preliminary comparison of the QUBO formulation with the QuarDP formulation on  Dirac-3. In addition to these comparisons, we also compared both the formulations' (QuarDP and QUBO) results with those of the CP-SAT solver from ORTOOLS.

As mentioned in Section~\ref{sec:ec_dirac-3} Dirac-3 allows you to set a specific number of discrete levels for each individual variable, up to a total of 954 levels collectively, for integer problems by interpreting multiple photons as a single discrete level.  (This mode is independent of the parameter $R$.) We ran our QUBO problems with two discrete levels (0 or 1) per variable and our QuarDP problems with eight discrete levels (0 to 7) per variable; both using the relaxation schedule 1 setting on the device. The QuarDP formulation involved fewer variables than the QUBO formulation, because the process of breaking a quartic function meant for integer/continuous variables into an equivalent quadratic one meant for binary variables adds a lot of auxiliary variables.
The CP-SAT solver in ORTOOLS is designed to solve integer programming problems. The NMF problem can be reduced to an integer optimization problem where the goal is to minimize the error in the objective function while adhering to the constraints. We do this in CP-SAT by first defining the variables and the range of integer values they can take. These include primary variables, such as the ones to represent $W$ and $H$, but also auxiliary variables like those used for keeping track of intermediate products, summations and reconstruction errors. Equality constraints are added between the appropriate primary and auxiliary variables. Based on how the matrix reconstruction errors were defined, we get the final objective function to be minimized. For this work, we implemented and tested the following two objective functions:
\begin{align}
    \argmin_{W,H}\sum_{i=1}^{n} \sum_{j=1}^{m}|V_{ij} - \sum_{k=1}^{p} W_{ik}H_{kj}|\label{eq:abs_diff} \\
    \argmin_{W,H}\sum_{i=1}^{n} \sum_{j=1}^{m}(V_{ij} - \sum_{k=1}^{p} W_{ik}H_{kj})^2\label{eq:sq_diff}
\end{align}
Both objective functions are based on the differences between $V_{ij}$ and $\sum_{k}W_{ik}H_{kj}$. Equation~\ref{eq:abs_diff} measures error as the sum of absolute differences and Equation~\ref{eq:sq_diff} measures error as the sum of squared differences; the latter being equivalent to the QuarDP objective function Equation~\ref{eq:nmf_frobnorm_sq}).
With objective functions based on 
Equations~\ref{eq:abs_diff}) and \ref{eq:sq_diff}, we ran CP-SAT ten times per test case and restricted it to a single processor on the laptop described before, but set to performance mode, a power setting that maximizes CPU performance.
This configuration made for a more equitable comparison with the Dirac-3 machine. The time limit for each CP-SAT run was set to the time Dirac-3 required to solve the same test case for QuarDP formulations.  Note that while we cannot at this time explicitly set a time limit for Dirac-3's runs, we can measure how long a run took once it is completed. Finally, the best of the ten runs, in terms of smallest relative error, is taken and stored.  This stored value was then compared with the best run from QUBO and QuarDP formulations.
\subsubsection{Results and Discussion}\label{sec:exp3_results}
\begin{figure}[htbp]
    \centering
    \subfloat[]{
        \includegraphics[width=8cm, height=6cm]{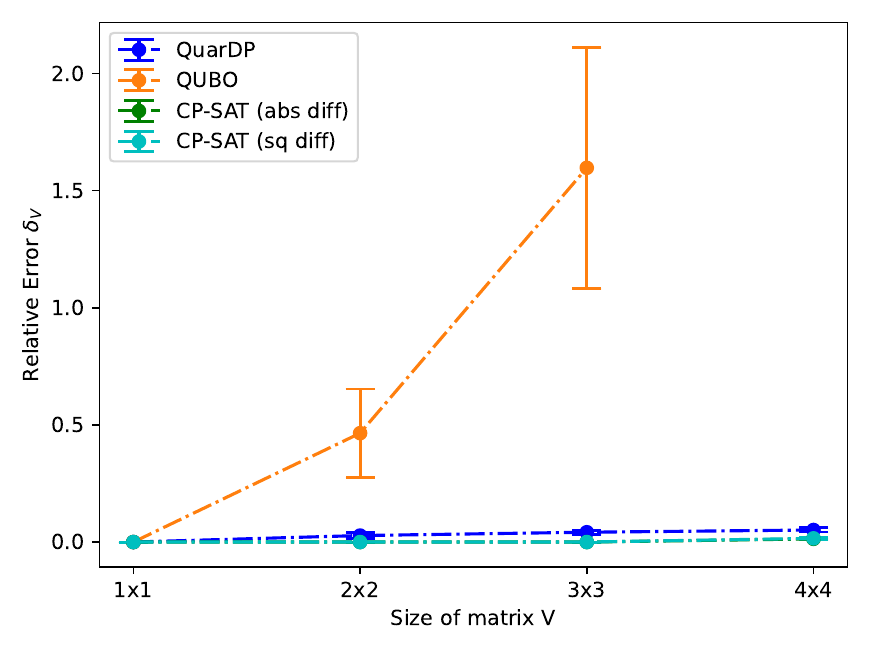}
        \label{subfig:exp3_rel_error}
    }
    \hfill
    \subfloat[]{
        \includegraphics[width=8cm, height=6cm]{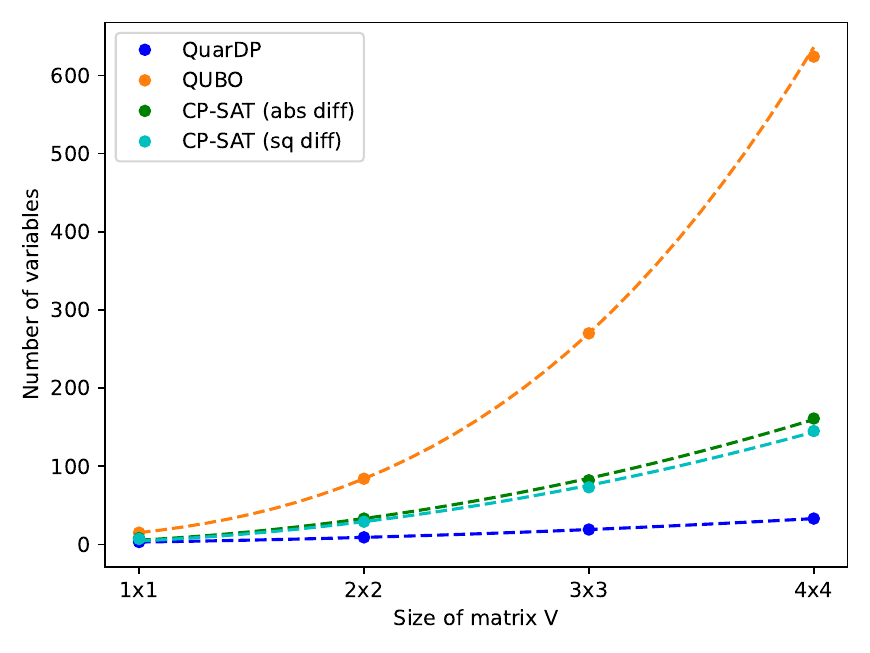}
        \label{subfig:exp3_numvar}
    }
    \caption{(a) Median relative errors $\delta_{V}$
 and (b) number of variables needed per solver/formulation type in Experiment III, shown as functions of matrix size. In (a), the $4\times4$ data point is missing for QUBO because Dirac-3 was not able to accommodate the problem size. Curves are fitted for the data in (b) but not in (a)
 .}
    \label{fig:exp3}
\end{figure}

Figure~\ref{subfig:exp3_rel_error} shows the growth of relative errors as the size of the matrix $V$ increases. Each y-value is the median of the result from 100 test cases per size of $V$ and the error bars measure median absolute deviation). The number of variables each solver/formulation needs per test size is shown in Figure~\ref{subfig:exp3_numvar}. The latter is curve-fitted to its data points, Cubic for QUBO, Quadratic for QuarDP and CP-SAT
However, the former is not, owing to (i) the absence of a $4\times4$ data point for the QUBO formulation (explained below) and (ii) CP-SAT consistently achieving zero relative error across multiple matrices, making curve-fitting even more inaccurate.
From Figure~\ref{subfig:exp3_rel_error}, we see that Dirac-3, with the QUBO formulation, performs the worst of all the different solvers. This can be attributed mainly to the fact that it has the largest number of variables to optimize; which is in turn a byproduct of the fact that it needs a lot of auxiliary variables to make the original quartic formulation into a quadratic one (as explained in
Section~\ref{sec:quartic_2_quadratic}). In fact, the test cases for $4 \times 4$ matrices were not able to fit on the machine because the total number of variables exceed the number of discrete levels for integer/binary optimization.  It is possible that other QUBO solvers perform better than Dirac-3 for this particular problem type, but that is outside the scope of this particular study.

We go on to observe is that both CP-SAT variants performed better than the QuarDP formulation on Dirac-3 (albeit marginally) even though they need more variables than Dirac-3 for the same matrix size. While we do not have conclusive proof on why this is the case, we have an explanation based on the results of Experiment III and Experiment IV to be described in the following section.

First, CP-SAT is a mature, sophisticated hybrid SAT-based solver enhanced with constraint propagation and linear programming (LP) relaxations \cite{cpsatlp}. In contrast, Dirac-3 is a nascent hardware-based solver, which is still in early stages of development \cite{Nguyen_2025}, and as such lacks the algorithmic refinements and optimizations that CP-SAT has accumulated over years of development.

Second, the matrix $V$ in the test cases for Experiment III had at least one pair of exact factors $W,H$. 
This makes it easy for a solver with SAT solving capability to halt instantly when a perfect solution is reached, that is, when $V=WH$. The Dirac-3, on the other hand, works on the principle of energy-based minimization in a stochastic manner, and cannot check for satisfiability conditions in the middle of its computation. Furthermore, while the optimal answer (the one that satisfies $V=WH$ in this case) should have the highest probability of being retrieved (lowest energy), there are often many `good' answers with a reasonable chance of being measured. As a consequence, the collective probability of measuring one of these good answers is generally higher than that of measuring the single best answer. There is some evidence to suggest that similar energy-based optimizers are better suited for MAXSAT problems, where the goal is to maximize the number of clauses satisfied,
rather than regular SAT problems \cite{Aadit_2022}.

\subsection{Experiment IV}\label{sec:exp4}
\subsubsection{Setup}\label{sec:exp4_setup}
In this final experiment, based on the lessons from Experiment III, we compared the QuarDP formulation running on Dirac-3 with CP-SAT with objective functions Equation~\ref{eq:abs_diff}) and Equation~\ref{eq:sq_diff}, but for rectangular matrices $V\in \mathbb{Z}^{4\times 8}$. Inspired by Experiment II, we generate two sets of test cases:
\begin{enumerate}
    \item Test set A: One hundred test cases where, for each test case, the matrix $V\in \{i\in\mathbb{Z}| 0\leq i \leq7\}^{4\times8}$ is generated using the {\tt randint()} function in numpy.
    \item Test set B: One hundred test cases where, for each test case, we first generate factor matrices $W\in S^{4\times3}$ and $H\in S^{3\times8}$ where $S = \{i\in\mathbb{Z}| 0\leq i \leq7\}$ (both generated using the {\tt randint()} function from the {\tt numpy} package). The matrix $V$ is then generated by multiplying $W$ and $H$.
\end{enumerate}
As in experiment II, test cases in set B has at least one pair of factor matrices $W,H$, but the same condition is not guaranteed for test cases in set A. These two sets will help us compare and contrast the behavior of Dirac-3 and CP-SAT.

The previous experiment compared the best runs from each solver/formulation. Here in Experiment IV, we compare the median runs (run with $\delta_{V}$ closest to the median) from each solver/formulation, providing a more representative performance comparison.  As before, we used the relaxation schedule 1 setting to run the test cases on Dirac-3, and CP-SAT's time limit was then set to match the run time of Dirac-3 for every test case.
Each CP-SAT run was done with the same laptop in performance mode, but now in the following two configurations: (i) one where the computation was restricted to a single processor (serial processing) and (ii) the other where CP-SAT was allowed to use all 12 processors in our Intel i5-1335U chip (parallel processing). Although serial processing provides a more direct comparison between Dirac-3 and CP-SAT, parallel processing is generally how CP-SAT is used in practice. The parallel version of CP-SAT not only executes multiple worker threads of different search strategies at once, but is also able to share information like learned clauses and bounds between each of its threads \cite{cpsatlp}.

\subsubsection{Results and Discussion}\label{sec:exp4_results}
\begin{figure*}[tbhp]
\centering
\subfloat[]{\includegraphics[width=8cm, height=6cm]{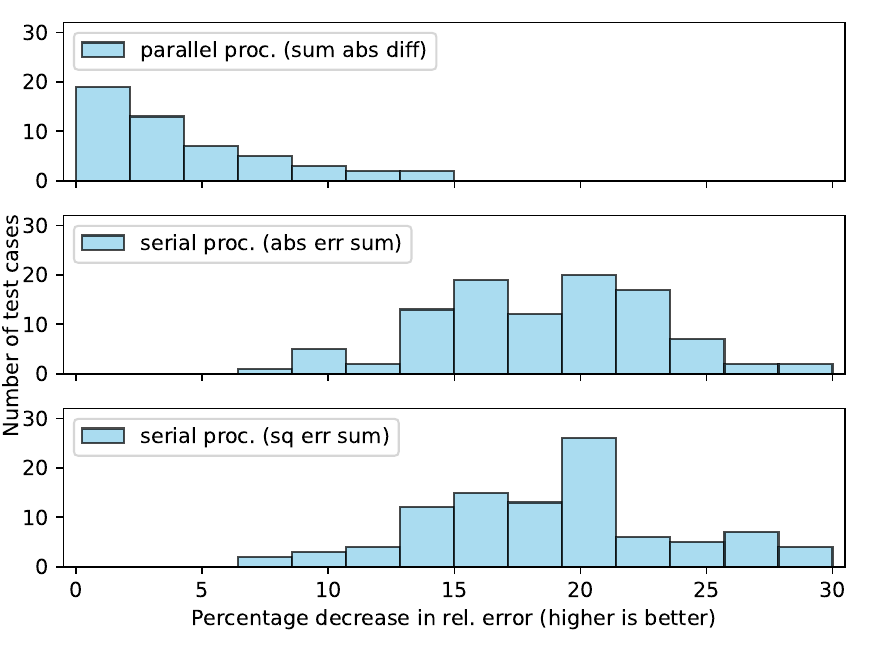}%
\label{subfig:hist_exp4_setA}}
\hfil
\subfloat[]{\includegraphics[width=8cm, height=6cm]{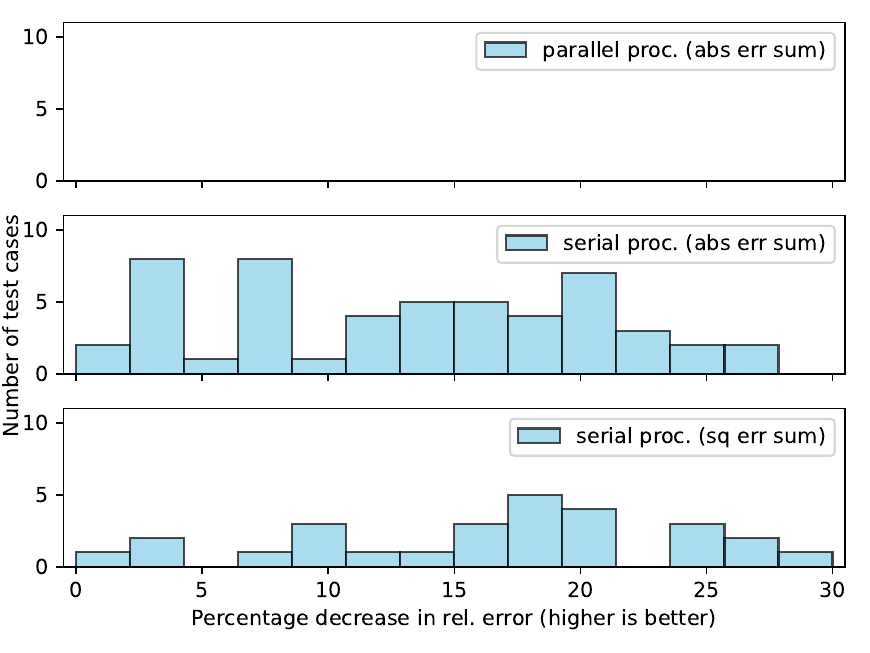}%
\label{subfig:hist_exp4_setB}}
\caption{Histograms for the test cases where Dirac-3 (with QuarDP formulation) outperformed the CP-SAT solver for (a) test set A and (b) test set B respectively as a measure of the percentage improvement in relative error $\varDelta\delta_{V}$.}

Each individual histogram shows the percentage improvement against a particular configuration of CP-SAT (processing type and objective function). Though no advantage was found against parallel CP-SAT for test set B, a blank histogram is plotted to maintain the symmetry between the two sub-figures. Note that both axes in the sub-figures are scaled differently, to appropriately display the range of results for each test set. 
\label{fig:hist_exp4}
\end{figure*}

Unlike Experiment III, Dirac-3 performs better (in terms of the relative error $\delta_V$) than CP-SAT in various scenarios, at least against the serial processing configuration. Dirac-3 performed especially well on test set A: problems that most likely do not have at least one pair of factors $W,H$ such that $WH=V$ without error.
For this experiment, we shall once again calculate the p-values using Equation~\ref{eq:p-val} for our alternative hypothesis of Dirac-3 being better than CP-SAT.

Table~\ref{tab:which_better_exp4_setA} shows that Dirac-3 outperformed CP-SAT in every problem instance when the latter was restricted to one processor. When CP-SAT was allowed to use all 12 available processors, that advantage went down to about half of the test cases for the objective function based on the sum of absolute differences as in Equation~\ref{eq:abs_diff}.  Furthermore, no advantages were observed when comparing Dirac-3 against the other objective function (sum of squared differences Equation~\ref{eq:sq_diff}). In fact, the calculated p-values fail reach the level needed to reject the null hypothesis for parallel CP-SAT for both test sets.

When it comes to Table~ \ref{tab:which_better_exp4_setB} (test set B), we can see that Dirac-3 outperforms single-processor CP-SAT on fewer instances. In addition, there were no cases where Dirac-3 performed better than CP-SAT in its parallel processing configuration. Here, the only statistically significant result in support of our alternative hypothesis is against CP-SAT running on a single processor with Equation~\ref{eq:abs_diff} as the objective function.

\begin{table}[htbp]
    \centering
\begin{tabular}{|c|c|c|c|c|}
\hline
\multirow{2}{*}{\diagbox{Method}{\makecell{Obj. func. \&\\ CPU mode}}} & \multicolumn{2}{c|}{Abs. diff.} & \multicolumn{2}{c|}{Sq. diff.} \\
\cline{2-5}
 & \makecell{serial\\proc.} & \makecell{parallel.\\proc.} & \makecell{serial\\proc.} &\makecell{parallel.\\proc.}\\
\hline
Dirac-3 better count& 100 & 51 & 100 & 0\\
\hline
CP-SAT better count& 0 & 49 & 0 & 100\\
\hline
p-value (Dirac-3 better) & 7.8e-31 & 0.46 & 7.8e-31 & 1\\
\hline
\end{tabular}
    \caption{Performance comparison by lower relative error: Dirac-3 vs. CP-SAT methods (test set A)}
    \label{tab:which_better_exp4_setA}
\end{table}

\begin{table}[H]
    \centering
\begin{tabular}{|c|c|c|c|c|}
\hline
\multirow{2}{*}{\diagbox{Method}{\makecell{Obj. func. \&\\ CPU mode}}} & \multicolumn{2}{c|}{Abs. diff.} & \multicolumn{2}{c|}{Sq. diff.} \\
\cline{2-5}
 & \makecell{serial\\proc.} & \makecell{parallel.\\proc.} & \makecell{serial\\proc.} &\makecell{parallel.\\proc.}\\
\hline
Dirac-3 better count& 59 & 0 & 41 & 0\\
\hline
CP-SAT better count& 41 & 100 & 59 & 100\\
\hline
p-value (Dirac-3 better) & 0.04 & 1 & 0.46 & 1\\
\hline
\end{tabular}
    \caption{Performance comparison by lower relative error: Dirac-3 vs. CP-SAT methods (test set B)}
    \label{tab:which_better_exp4_setB}
\end{table}

For the cases where an advantage was achieved, we measure it using the percentage decrease in relative error (against CP-SAT) metric:
\begin{align}
    \varDelta\delta_{V} = \frac{\delta_{V}^{(CP-SAT)} - \delta_{V}^{(Dirac-3)}}{\delta_{V}^{(Cp-SAT)}} \times 100\label{eq:rel_error_percentage_cpsat}
\end{align}
Here, Equation~\ref{eq:rel_error_percentage_cpsat}is a variant of Equation~\ref{eq:rel_error_percentage} where sklearn is replaced by CP-SAT and fusion is replaced by Dirac-3. Using this metric, we plot histograms to compare the Dirac-3 results with different CP-SAT variants. Figure~\ref{fig:hist_exp4} shows us the general distribution for cases where an advantage was observed. For CP-SAT running on a single processor, considering both the objective functions, the median improvement was $\approx15\%$ (for test set A) and $\approx18\%$ (for test set B). Dirac-3 outperformed CP-SAT in half of the parallel processing test cases with the sum of absolute differences objective function (only for test set A), though the median advantage was only approximately 4\%.

In the previous experiment, Dirac-3 was not able to outperform CP-SAT at all. However, here in Experiment IV, Dirac-3 was able to do so in plenty of cases. The major findings from this experiment and their possible explanations are as follows:
\begin{enumerate}
    \item The number of variables required by CP-SAT for this experiment's problem size exceeded those needed for the largest problem in the previous experiment.  Specifically, CP-SAT required 261 and 229 variables for 
    Equations~\ref{eq:abs_diff} 
    and~\ref{eq:sq_diff}) respectively, compared to 161 and 145 variables in Experiment III.  In contrast, Dirac-3 required only 37 variables for these problems, an increase of just 5 variables from Experiment III. This rapid growth in CP-SAT's variable count may have caused it to reach a crossover point where Dirac-3's performance became competitive.
    \item On closer inspection, we find that Dirac-3 was more competitive against CP-SAT for test set A than for test set B. This could simply be due to the nature of the problems. Indeed, all the test cases in experiment III and test set B have at least one pair of factor matrices $W,H$ such that $V=WH$. This makes it easier for CP-SAT to halt as soon as the equality condition is satisfied, or to store/remember a solution that satisfies the equality condition. As we explained in Section~\ref{sec:exp3_results}, a stochastic energy-based optimizer does not operate in that way.
    \item Related to the above, we believe one reason CP-SAT could have underperformed for test set A (primarily in serial processing) could have been the lack of exact factor matrices $W,H$. While CP-SAT is a competent optimizer, we believe this type of a problem may be a better match for a solver like the Dirac-3.
    \item Collectively, wherever CP-SAT performed better, it did so more with the objective function based on Equation~\ref{eq:sq_diff} than for the function based on Equation~\ref{eq:abs_diff}. This may be the case because we measure $\delta_V$ and $\varDelta\delta_V$ with the Frobenius norm, which is very similar to Equation~\ref{eq:sq_diff}. In other words, the objective function for Equation~\ref{eq:sq_diff}) is similar in structure to the metric used to measure its performance.
    \item CP-SAT's parallel processing is highly effective, even when running on an ordinary laptop. While it may be tempting to make conclusions about entropy computing being uncompetitive against CP-SAT based on this observation, it should be noted that the current implementation uses a hybrid photonic-electronic architecture with a computational bottleneck in the measurement-feedback loop, specifically in the tomography process used to determine variable values for candidate solutions. This bottleneck limits the iteration rate of the optimization. A more equitable comparison for parallel CP-SAT would be against an all optical implementation of entropy computing, or against an entropy computer with multiple optimization processes running at once. However, parallel CP-SAT has other advantages, such as the ability to share information among its parallel worker threads. A truly competitive NVN computer may also need some mechanism of information sharing between parallel optimization processes, such as in parallel tempering \cite{Swendsen_1986,geyer1991markov,Aadit_2023}.
    \item While parallel CP-SAT performs well, scaling to larger problems would require more powerful CPUs with correspondingly higher energy consumption. Entropy computers may achieve better performance-per-watt scaling, 
    depending on the measure of performance used, as photonic architectures typically have lower power requirements\cite{Xu_2025, Fayza_2025}. Further studies are needed to confirm this.
\end{enumerate}
\section{Future work}
\label{sec:futureWork}
Based on the results of our four experiments, we believe that the community would benefit from the following five directions of research work.

First, an extension of this work to include regularization terms for the matrices $W$ and $H$ can increase its utility and practicality. This can be done by including the regularization parameter in the optimization's objective function \cite{Li_2018,Desu_2021}.

Second, a hybrid method for larger problems that uses conventional CPUs in conjunction with a device like Dirac-3 could be useful. One way this method may be implemented is by (i) generating initial factor matrices $W_{0},H_{0}$ (by using a technique like NNSVD) then iteratively, (ii) creating an objective function by designating a limited number of elements (that can fit on Dirac-3) in $W$ and $H$ as variables and the rest as constants (fixing their values from the previous iterations), (iii) solving the objective function and getting the values for the designated elements, and (iv) repeating steps two and three until a convergence point is reached.  (This process is inspired by the BCD method~\ref{alg:BCD}.)

Third, it would be interesting to see if an all-optical implementation of an entropy computer can outperform parallel CP-SAT, at least for a certain number of processors. Furthermore, it may beneficial to compare and contrast results from other QUBO solvers such as quantum annealers, P-bit machines, coherent Ising machines, etc., against conventional solvers (like CP-SAT) for this problem type.

Fourth, given the fundamental architectural differences between photonic and digital systems, analyzing energy requirements and their scaling across problem sizes may reveal efficiency advantages beyond what runtime comparisons alone can show. This would require measuring power consumption from laser pumps, FPGAs, and other system components and compare it to CPU baselines.

Finally, based on the strong preliminary results for integer NMFs. It may be interesting to find application domains and variants of the integer NMF problems where Dirac-3 and other NVN optimizers have a clearer advantage.

\section{Conclusion}
\label{sec:conclusion}
In this work, we proposed formulations to solve NMF problems using energy-based optimization-centric NVN architecture computers, with a focus on entropy computing. Our formulations attempt to simultaneously obtain both factor matrices $W$ and $H$. We found our quartic formulation QuarDP for integer/continuous values to be better than the binary QUBO model approach. For real-valued NMF problems, we observed that while the Dirac-3 entropy computer was not able to directly outperform sklearn's NMF procedure, a fusion approach that used Dirac-3's solutions as the initial matrices $W_0,H_0$ in the sklearn's NMF procedure outperformed the standard version of the same, in terms of relative error. In particular, a notable improvement was seen for test cases where the NMF problems had exact factors. For integer NMF problems, when comparing the CP-SAT integer programming solver running on a single processor against Dirac-3 for the same run time per problem, we were able to observe an advantage in a majority of the test cases. However, the results from the parallel version of CP-SAT were generally better than Dirac-3's. We believe that one of the primary reasons for this could be a computational bottleneck due to Dirac-3's hybrid architecture; and we would expect an all optical implementation of entropy computing to yield better results.

\section*{Acknowledgments}
The authors thank Drew Rotunno and Lac Nguyen from Quantum Computing Inc (QCi) for making this collaboration possible.

\section*{Code and data availability}
The code and data associated with this paper can be found at \href{https://github.com/aborle1/NMF_NVN/}{https://github.com/aborle1/NMF\_NVN/}.

\section*{AI declaration statement}
In this work, Anthropic's Claude\textsuperscript{TM} Sonnet 4.5 large language model (LLM) was used for grammar checking and sentence restructuring, as well as to generate starter code for the CP-SAT solver (which was subsequently verified by the authors). Furthermore, the LLM was also used to fit curves for the data points in Figure \ref{subfig:exp3_numvar}.

{\appendix[An Overview of Entropy Computing and Dirac-3]\label{appendix:entropy}
In its purest conceptualization, entropy computing is a computational paradigm where the solution candidates (i.e values for our variables) are encoded as photonic quantum states in terms of photons and their modes. The initial photonic state can be considered as a random candidate solution for the optimization problem in question. The objective function of the problem is encoded as a Hamiltonian that determines the differential loss rate for the photonic state in question. A candidate solution that has a higher cost will have its photonic state attenuated more than a candidate solution with a lower cost; this happens with the help of a non-linear loss medium. Finally, the surviving signal is amplified. However, the amplifier also induces quantum fluctuations in the state proportional to the amount of attenuation. This translates to a larger change in the photonic states that have a higher cost and little or no change to the states with lower costs. In other words, bad candidate solutions have their values changed more than good candidate solutions. This loop is iterated multiple times with the help of a relaxation schedule, using the device's control parameters, to suppress solutions with higher costs and evolve those with lower costs.

Although an all-optical implementation of entropy computing would have been preferred, the technology, at the time of this work, is implemented using a hybrid photonic-electronic approach in the Dirac-3 entropy computer \cite{Nguyen_2025}. Figure~\ref{fig:dirac_3} shows the various components used in Dirac-3. In essence, certain tasks that are currently hard to implement with optics are realized using electronic components. The most important among these components are field programmable gate arrays (FPGAs) that calculates the loss rate of a candidate solution for the given problem Hamiltonian. But in order to use an FPGA, the photonic state needs to be converted into electrical signals. This requires us to do a quantum tomography of the state by measuring our prepared photonic state multiple times (shots) and using a single photon detector for different time bins to get a histogram of the distribution of photons per qudit, using a single photon detector and a time correlated photon counter. The shot noise present while taking the measurements becomes the source for inducing noise/fluctuations, which manifest as randomness in our current candidate solution. Once the FPGA ascertains the quality of our current candidate solution and calculates the loss rate, the solution is encoded again as a photonic state through a digital to analog converter (DAC) and an electro-optical modulator that also implements the loss mechanism. For further details on this, we recommend the original paper by Nguyen et. al \cite{Nguyen_2025}.
\begin{figure}[htbp]
\centering
\includegraphics[width=8.5cm, height=6cm]{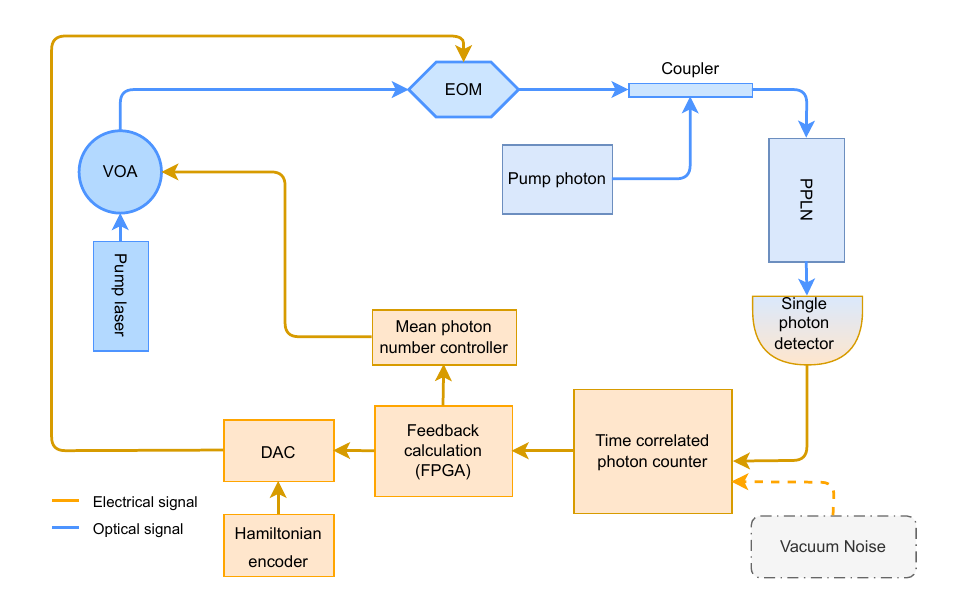}
\caption{Block diagram of the Dirac-3 entropy computer. Here, variable optical attenuators (VOA) help in generating the optical signal along with a pump laser and  a mean photon number controller. This feeds into the electro optical modulator (EOM), which creates the temporal wavefunction and implements the loss mechanism with the help of a digital-to-analog converter (DAC). A periodically-poled lithium niobate (PPLN) waveguide does the frequency up-conversion  of the photons in the optical signal for better detection. Finally, measurements are taken with the help of a single photon detector and a time-correlated photon counter (for multiple shots) and a candidate solution is created from the tomography. A field programmable gate array (FPGA) then calculates the new loss rate based on the candidate solution measured in the context of the problem Hamiltonian, and the cycle repeats until a termination criteria, such as a timeout or convergence to a candidate solution, is satisfied. 
\label{fig:dirac_3}}
\end{figure}


 

\bibliographystyle{IEEEtran}
\bibliography{references}

\newpage

 





\end{document}